\documentclass{article}
\usepackage{spconf,amsmath,epsfig}
\usepackage{graphicx}
\usepackage{subfigure}
\usepackage{colortbl}
\usepackage[table*]{xcolor}
\usepackage{algorithm}
\usepackage{algorithmic}
\usepackage{multirow}
\usepackage{multicol}
\usepackage{amssymb}
\usepackage{amsmath}
\usepackage{cite}
\usepackage{caption}
\usepackage{multirow}
\usepackage{booktabs}

\title{SEMI-SUPERVISED COMPLEX-VALUED GAN FOR POLARIMETRIC SAR IMAGE CLASSIFICATION}
\name{Qigong Sun, Xiufang Li, Lingling Li, Xu Liu, Fang Liu, Licheng Jiao
\thanks{This work was supported in part by the State Key Program of National Natural Science of China (No. 61836009, No.  91438201 and No. 91438103),
the National Natural Science Foundation of China (No. 61871310, No. 61876220).}}

\address{Key Laboratory of Intelligent Perception and Image Understanding of Ministry of Education,\\
International Research Center for Intelligent Perception and Computation,\\
Joint International Research Laboratory of Intelligent Perception and Computation,\\
School of Artificial Intelligence, Xidian University, Xi’an, Shaanxi Province 710071, China}
\begin{document}
%
\maketitle
\begin{abstract}
Polarimetric synthetic aperture radar (PolSAR) images are widely used in disaster detection
and military reconnaissance and so on.
However, their interpretation faces some challenges, e.g., deficiency of labeled data, inadequate utilization of data information and so on.
In this paper, a complex-valued generative adversarial network (GAN) is proposed for the first time to address these issues.
The complex number form of model complies with the physical mechanism of PolSAR data and in favor of utilizing and retaining amplitude and phase information of PolSAR data. GAN architecture and semi-supervised learning are combined to handle deficiency of labeled data.
GAN expands training data and semi-supervised learning is used to train network with generated, labeled and unlabeled data.
Experimental results on two benchmark data sets
show that our model outperforms existing state-of-the-art models, especially for conditions with fewer labeled data.
\end{abstract}
\section{Introduction}
\label{sec:intro}
\vspace{-0.3cm}
Many researches have
been done on PolSAR image classification, and breakthrough benefits from the development and application of
deep convolutional neural networks(DCNN) \cite{krizhevsky2012imagenet}.
As we all know, PolSAR data are usually expressed by coherent matrices or covariance matrices which contain amplitude and phase information in complex number form.
However, a general real-valued DNN loses significant phase information when it is applied to interpret PolSAR data directly.
\cite{zhou2016polarimetric} converts a complex-valued coherent or covariance matrix into a normalized 6-D real-valued vector for PolSAR data classification, while ignoring important phase information.
Different from direct conversion of complex number into a real number, some other strategies are introduced. Besides the coherency matrix extended to the rotation domain,
Chen \textsl{et al}. \cite{chen2018polsar} also take the null angle and roll-invariant polarimetric features as input to extract ample polarimetric features. Liu \textsl{et al}. \cite{Liu2018Polarimetric} propose a novel polarimetric scattering coding method for gaining more polarimetric features in classification. However, their operations are all in the real number domain.

Instead, in order to make full use of PolSAR data information,
some complex-valued DNN models are proposed.
Inspired by the application of complex-valued convolutional neural network (CV-CNN)\cite{guberman2016complex},
Zhang \textsl{et al}\cite{Zhang2017Complex} proposed the application of CV-CNN on PolSAR data classification and obtained a great success.
This is the beginning of CV-CNN to classify PolSAR data.
Besides retaining information, CV-CNN has the strengths of faster learning and converenge\cite{nitta2002critical}.
In addition, deep learning is a data-driven approach. However, the labeled samples are
extremely deficient in PolSAR data.
Thus, unsupervised or semi-supervised networks are used for the classification of PolSAR data, for example, deep convolutional autoencoder\cite{Geng2015High}.
Meanwhile, GAN\cite{goodfellow2014generative} is able to expand data. It can learn the potential distribution of actual data and generate fake data that has the same distribution with actual data.
With the successful application in many fields (the generation of natural images\cite{radford2015unsupervised} and Neural Dialogue\cite{li2017adversarial} and so on), the GAN architecture has received increasing attention in recent years. In order to further solve the deficiency of
labeled data, it is advisable to combine GAN architecture
and semi-supervised learning. Therefore, in this paper, we propose a complex-valued GAN framework.

Our novel model has three advantages: 1) The complex-valued neural network complies with the physical mechanism of the complex numbers, and it can retain amplitude and phase information of PolSAR data; 2) GAN extended to complex number field can expand PolSAR samples, which have similar distribution with actual samples. Increased samples can improve the classification performance of PolSAR data.
3) Besides labeled data, unlabeled data are also used to update model parameters by semi-supervised learning and improve network performance to a certain extent.
\vspace{-0.2cm}
\section{SEMI-SUPERVISED COMPLEX-VALUED GAN}
\label{sec:semi-supervised gan}
\vspace{-0.2cm}
\subsection{Network Architecture}
\label{sec:Architecture}
\vspace{-0.2cm}

The data generated by general real-valued GAN is different from PolSAR data in feature and distribution. Therefore, we extend real-valued GAN to the complex number domain and propose a complex-valued GAN.
Figure 1 illustrates the framework of our model,
and it is composed by Complex-valued Generator and Complex-valued Discriminator.
This framework consists of complex-valued full connection,
complex-valued deconvolution, complex-valued convolution,
complex-valued activation function and complex-valued batch
normalization, which are represented by "CFC", "CDeConv",
"CConv", "CA" and "CBN", respectively. In addition, a complex-valued network also makes full use of the
amplitude and phase features of PolSAR data.
\begin{figure}[!htb]
\centering
\setlength{\abovecaptionskip}{-0.0cm}
\setlength{\belowcaptionskip}{-0.4cm}
\includegraphics[width=3.4in]{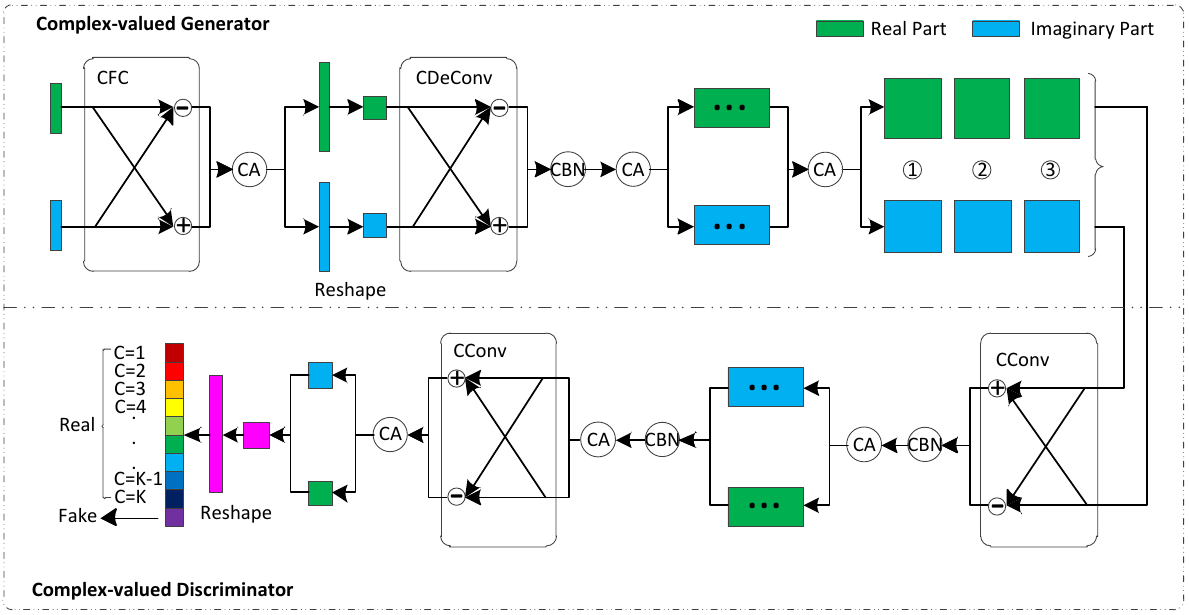}
\caption{\footnotesize{The framework of semi-supervised complex-valued GAN for image classification.
$\ominus$ denotes minus arguments in element-wise and $\oplus$ denotes adds arguments in element-wise.}}
\label{fig_University}
\end{figure}

In the Complex-valued Generator, after a serious of complex-valued operations,
two randomly generated vectors shown as the green block
and blue block are translated into a complex-valued matrix,
which has the same shape and distribution with PolSAR data.
In the Complex-valued Discriminator,
we use complex-valued operations to
extract complete complex-valued features, which are in the form of a pair.
Then we concatenate the real part and imaginary part of the last
feature to the real domain for final classification.
In the training processing, generated fake data, labeled and unlabeled actual data are used to
alternately train this complex-valued
GAN by semi-supervised learning, and until the network can effectively identify the authenticity of input data and achieve
correct classification.

\vspace{-0.3cm}
\subsection{Complex-Valued Operation Mask}
\label{sec:Mask}
\vspace{-0.2cm}

For simplifying the calculation, we choose the algebraic form to express a complex number.
In the algebraic form, the numbers in real part and imaginary part are real numbers with one dimension.
We use $z_1=a+ib$ and $z_2=c+id$ to denote two complex numbers,
the multiplication and addition are redefined as follows:
\vspace{-0.4cm}
\begin{small}
\begin{eqnarray}
z_1*z_2&=&(a+ib)*(c+id) \\\notag \
&=&(a*c-b*d)+i(a*d+b*d) \\
z_1\pm z_2&=&(a\pm c)+i(b\pm d)
\end{eqnarray}
\end{small}


To indicate the complex-valued operation mentioned in detail,
a complex-valued operation mask is proposed, as shown in Figure 2.
The green and the blue block represents the real and imaginary part, respectively.
This mask can make some complex number calculations,
whose input data ($IN\_r$, $IN\_i$), the
weight ($W\_r$, $W\_i$) and output data ($OUT\_r$, $OUT\_i$) are consisted of a real part and an imaginary part.
Therefore, this type of operation can be
decomposed to four traditional real operations, one addition
operation and one subtraction operation.
Each complex-valued operation in our network complies with this mask.
The same expression and physical mechanism of data and network parameters in favor of obtaining full data features used for classification.
\vspace{-0.4cm}
\begin{figure}[!htb]
\centering
\includegraphics[width=2in]{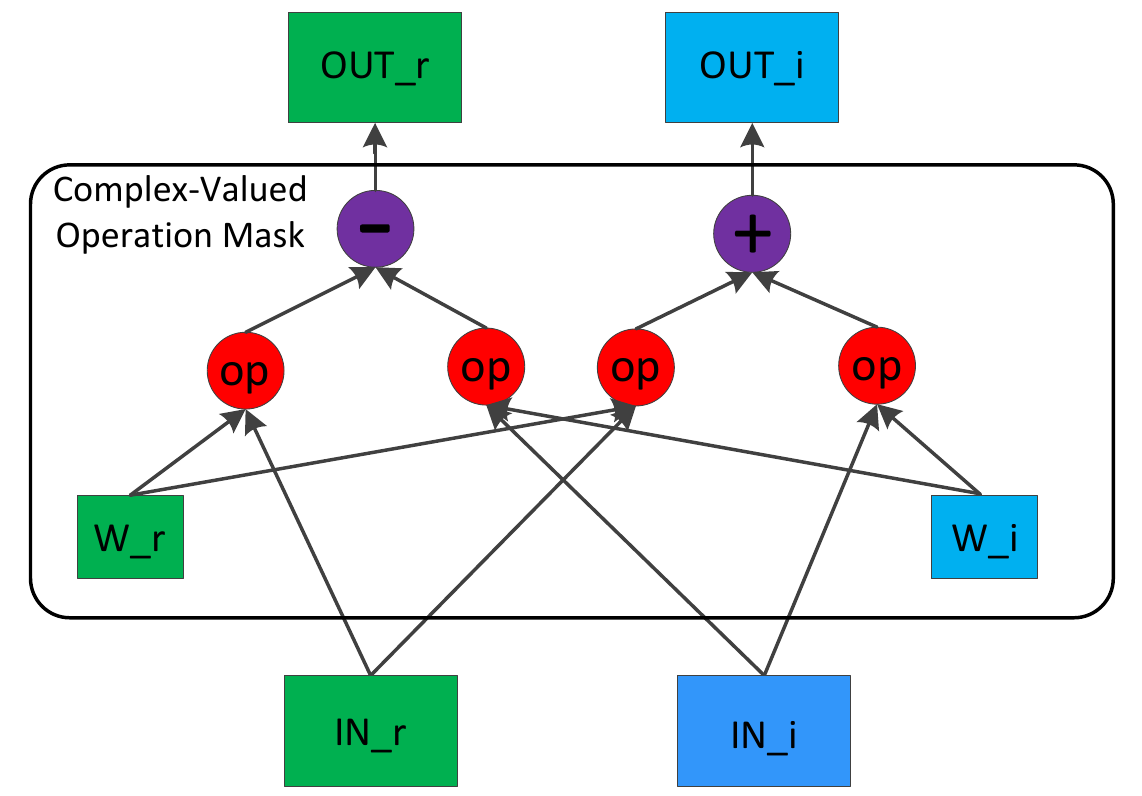}
\caption{\footnotesize{Complex-Valued Operation Mask.
The circular block denotes real-valued operations, the red circles are undetermined operations and the violet are explicit operations. "op" can be full connection, convolution or deconvolution.}}
\label{fig_University}
\end{figure}
\vspace{-0.4cm}
\vspace{-0.5cm}
\subsection{Complex-Valued Batch Normalization}
\label{sec:Batch Normalization}
\vspace{-0.2cm}

Batch normalization has been widely used in deep neural networks for unifying data and
accelerate convergence rate. In addition, complex-valued batch normalization can
stabilize the performance of GANs.
However, scanty training samples and less batch sizes restrict the effect of batch normalization.

In order to address this issue, a novel batch normalization is proposed in this paper.
The expectation and covariance matrices are replaced by
constantly updated average expectation and covariance matrices, so that they hold all sample information in training proceeding.
The following formulation shows the normalization of
the \emph{t}th batch x$_t$ :
\vspace{-0.3cm}
\begin{small}
\begin{eqnarray}
\hat{x_t}=(\bar{V}_t)^{-\tfrac{1}{2}}(x_t - \bar{\sigma}_t)
\end{eqnarray}
\end{small}
where $\bar{\sigma}_t$ and $\bar{V}_t$ represent the average expectation and covariance matrix from $t-m$ to $t$ batches, which is computed as follows:
\vspace{-0.4cm}
\begin{small}
\begin{eqnarray}
\bar{\sigma}_t &=& \frac{1}{m}\sum_{t-m}^{t}E\left [ x_t \right ]                              \\
\bar{V}_t &=&
\begin{bmatrix}
\bar{V}_{rr}^t & \bar{V}_{ri}^t\\
\bar{V}_{ir}^t & \bar{V}_{ii}^t
\end{bmatrix} \\ \notag \
&=&
\begin{bmatrix}
\frac{1}{m}\sum_{t-m}^{t}V_{rr}^t  & \frac{1}{m}\sum_{t-m}^{t}V_{ri}^t  \\
\frac{1}{m}\sum_{t-m}^{t}V_{ir}^t  & \frac{1}{m}\sum_{t-m}^{t}V_{ii}^t
\end{bmatrix}
\end{eqnarray}
\end{small}
where $m$ denotes the length of state remembered, and $\bar{V}_{ri}$ is equal to $\bar{V}_{ir}$.
The square root of a Matrix of 2 times 2
$\bar{V}_t$ is computed:
\vspace{-0.4cm}
\begin{small}
\begin{eqnarray}
\setlength{\abovedisplayskip}{1pt}
\setlength{\belowdisplayskip}{1pt}
S_t&=&(\bar{V}_{rr}^{t}\times \bar{V}_{ii}^{t} - \bar{V}_{ri}^{t} \times \bar{V}_{ri}^{t})^{\tfrac{1}{2}} \\
T_t&=&(\bar{V}_{rr}^{t}+ \bar{V}_{ii}^{t} + 2S_t)^{\tfrac{1}{2}} \\
\bar{V}_t^{-\tfrac{1}{2}}&=&
\begin{bmatrix}
\tfrac{(\bar{V}_{ii}^t + S_t)}{S_tT_t} & -\tfrac{\bar{V}_{ri}^t}{S_tT_t}\\
-\tfrac{\bar{V}_{ri}^t}{S_tT_t} & \tfrac{(\bar{V}_{rr}^t + S_t)}{S_tT_t}
\end{bmatrix}
\end{eqnarray}
\end{small}

\vspace{-0.2cm}
This operation can translate the data mean to 0 and variance to 1.
Ultimately, we use the following computing to denote complex-valued batch
normalization:
\begin{small}
\begin{equation}
\setlength{\abovedisplayskip}{1pt}
\setlength{\belowdisplayskip}{1pt}
BN(\hat{x}_t)= \gamma \hat{x}_t + \beta
\end{equation}
\end{small}
where $\gamma$ and $\beta$ are defined as two parameters to reconstruct the distribution.

\vspace{-0.5cm}
\subsection{Semi-Supervised Learning}
\label{sec:Learning}
\vspace{-0.2cm}

In this complex-valued GAN, for further utilizing features of unlabeled data,
we use semi-supervised learning to optimize network with a classifier of softmax.
The output of generator (G) is a $K+1$ dimensional vector $\{p_1,p_2,...,p_K,p_{K+1}\}$, where
from $p_{1}$ to $p_{K}$ are the probability of first K classes and $p_{K+1}$ is the
probability of input image being fake. In order to optimize
the generator (G) and discriminator (D), we define the loss
function as follows:
\vspace{-0.2cm}
\begin{small}
\begin{eqnarray}
L &=& L_{labeled} + L_{unlabeled} + L_{generated}            \\
L_{labeled}&=&-E \left [ logP(C|X_{real}, C<K+1) \right ]         \\
L_{unlabeled}&=&-E \left [ log\left [1 - P(C=K+1|X_{real}) \right ] \right ]      \\
L_{generated}&=&-E \left [ logP(C=K+1|X_{fake}) \right ]
\end{eqnarray}
\end{small}
where $L_{labeled}$, $L_{unlabeled}$ and $L_{generated}$ represent classification loss of
labeled samples, unlabeled samples, and generated samples,  respectively.
Therefore, classification losses of labeled and generated samples are easily acquired.
However, the classification loss of unlabeled samples is not easy to express because of
inexplicit ground truth. With this inevitable problem,
the output probability of softmax is operated as follows:
\vspace{-0.4cm}
\begin{equation}
\setlength{\abovedisplayskip}{0.5pt}
\setlength{\belowdisplayskip}{0.5pt}
p_{sum}=log\sum_{i=1}^{K}e^{(p-p_{max})} + p_{max}
\end{equation}
where $p_{max}$ denotes the max value in $p_i$ $(i<K+1)$, and
logistic regression as a binary classification is utilized.
When the output approaches 1,
the probability $p_{K+1} << p_{sum}$ accordingly,
the facticity of data is discriminated.
By this deduction, unlabeled data can also be used to update our network model.

\vspace{-0.5cm}
\section{EXPERIMENTS}
\label{sec:EXPERIMENTS}
\vspace{-0.2cm}

In our experiments, two benchmarks data sets of Flevoland and San Francisco are used.
In order to verify the effectiveness of our method, our model is compared with complex-valued convolutional
neural network (CV-CNN) and real-valued convolutional neural
network (RV-CNN), they have similar configurations with our Complex-valued Discriminator.
The overall accuracy (OA),
average accuracy (AA), and Kappa coefficient are used to
measure the performance of all the methods.

\renewcommand
\arraystretch{1.4}

\vspace{-0.5cm}
\subsection{Experiments on Standard Data Set}
\label{sec:Images Description}
\vspace{-0.2cm}
We use a coherent matrix T, which is a $3\times3$ conjugate symmetrical complex value matrix
and follows complex Wishart distribution, to express all information of the corresponding pixel on PolSAR images.
In Flevoland data, 0.2\%, 0.5\%, 0.8\%, 1.0\%, 1.2\%, 1.5\%, 1.8\%, 2.0\%, 3.0\%, 5.0\% labeled
data in each of 15 categories are randomly selected as training data,
and the remained labeled data for testing. In addition, 10\% unlabeled samples are used to
train our semi-supervised complex-valued GANs. In San Francisco data,
we randomly chose 10, 20, 30, 50, 80, 100, 120,150, 200, 300 labeled data in each of the 5 categories for training and 10\% data, no matter whether labeled, as actual samples.

\begin{figure}[!htb]
\centering
\setlength{\abovecaptionskip}{-0.0cm}
\setlength{\belowcaptionskip}{-0.6cm}
\includegraphics[width=3.4in,height=1.0in]{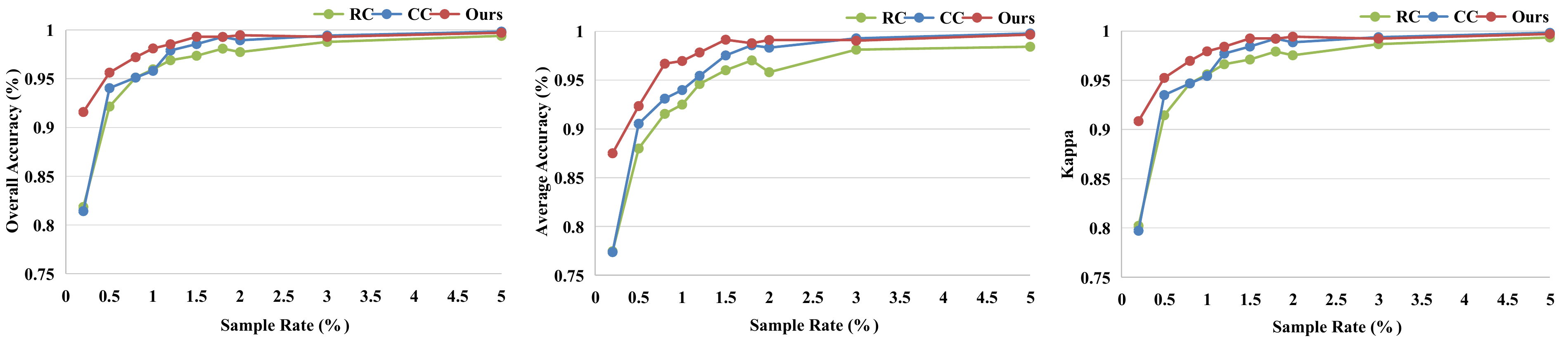}
\caption{\footnotesize{Flevoland OA, AA, and Kappa in different sample ratios.}}
\label{fig_University}
\end{figure}

\begin{figure}[!htb]
\centering
\setlength{\abovecaptionskip}{-0.0cm}
\setlength{\belowcaptionskip}{-0.4cm}
\includegraphics[width=3.4in,height=1.0in]{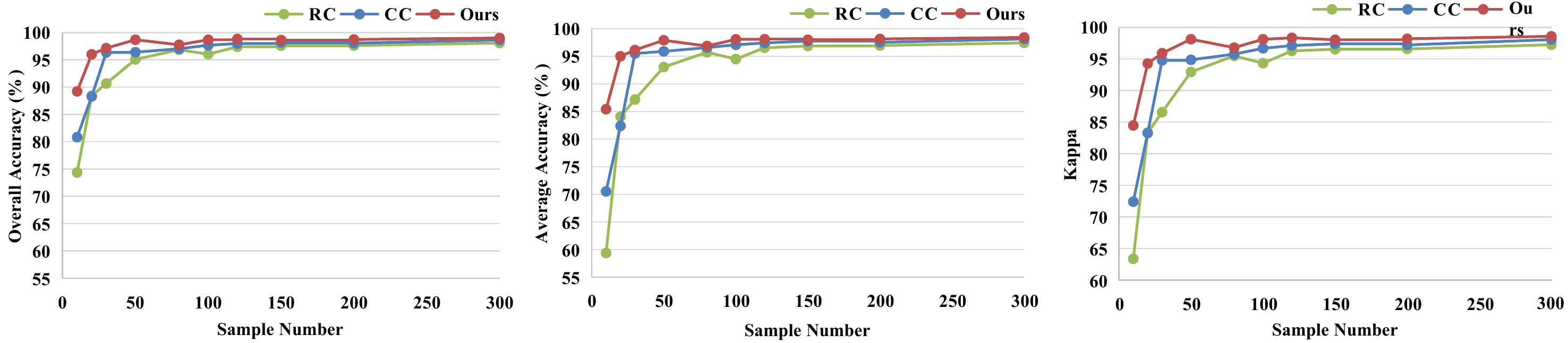}
\caption{\footnotesize{San Francisco OA, AA, and Kappa in different sample numbers.}}
\label{fig_University}
\end{figure}

\begin{table}[htbp]
\tiny
\centering
\setlength{\abovecaptionskip}{0.2cm}
\setlength{\belowcaptionskip}{-0.2cm}
\caption{\footnotesize{Classification accuracy(\%), OA(\%), AA(\%) and Kappa}}
\begin{tabular}{p{0.8cm} |p{0.45cm}|p{0.28cm}|p{0.28cm}|p{0.28cm}|p{0.28cm}|p{0.28cm}|p{0.28cm}|p{0.28cm}|p{0.28cm}|p{0.28cm}|p{0.28cm}|p{0.28cm}|p{0.28cm}|p{0.28cm}|p{0.28cm}|}
\hline
\hline
\multirow{8}{*}{Flevoland}& methods & 1 & 2 & 3& 4& 5 & 6 & 7& 8& 9\\
\cline{2-11}
 &RC& 87.18& 97.85& 95.56& \textbf{94.58}& 86.72& 93.96& 98.17& 98.89& 96.70\\
 &CC& 90.79& 98.39& 95.95& 89.71& 93.00& 93.21& 97.46& 99.24& 97.54\\
 &ours& \textbf{98.22}& \textbf{99.25}& \textbf{99.29}& 86.71& \textbf{95.40}& \textbf{95.27}& \textbf{99.85}& \textbf{99.85}& \textbf{98.59}\\
\cline{2-11}
 &methods& 10 & 11& 12& 13 & 14 & 15& OA& AA & Kappa\\
\cline{2-11}
 &RC& 94.88& 97.70& 83.45& 95.56& 99.00& 52.95& 95.12& 91.54& 94.68\\
 &CC& \textbf{98.02}& 97.01& 91.18& 90.48& 98.91& 65.57& 95.12& 93.10& 94.68\\
 &ours& 97.56& \textbf{97.76}& \textbf{96.07}& \textbf{99.06}& \textbf{100.0}& \textbf{87.38}& \textbf{97.21}& \textbf{96.68}& \textbf{96.97}\\
\hline
\multirow{4}{*}{San Francisco}& methods & 1 & 2 & 3& 4& 5& OA& AA & Kappa \\
\cline{2-10}
 &RC& 99.16& 86.86& 59.93& 19.29& 31.52&74.36 &59.35 &63.37 \\
 &CC& 99.07& 84.05& 53.81& \textbf{65.51}& 50.14&80.83 &70.51 &72.41\\
 &ours& \textbf{99.45}& \textbf{88.33}& \textbf{86.72}& 61.91& \textbf{90.61}&\textbf{89.23} &\textbf{85.41} &\textbf{84.48}\\

\hline
\hline
\end{tabular}
\end{table}

The parameters of all experiments in this paper
are set as follows: the patch size is $32 \times 32$, the learning rate is 0.0005, and the optimization method
is Adam with $\beta1=0.5$ and $\beta2=0.999$. Figure 3 and Figure 4 show the change of OA, AA,
and Kappa with the sample ratio in two data sets.
In Flevoland data, the results verified the superiority of our new network with less
labeled samples, and this law especially obvious when training samples less than 3.0\%.
This same advantage also is shown in San Francisco data,
especially if numbers of training data less than 50. In order to exhibit the contributions of our model on
each category, we list all test accuracy of Flevoland data with 0.8\% sampling ratio
and of San Francisco data with 10 labeled training samples in Table 1.
In Flevoland data, we can find that accuracies of different categories have generally improved especially for the fifteenth category, which
has the least training samples and achieves increase of 65.1\% and 33.17\% compare to the real-valued and complex-valued neural networks in accuracy, respectively.
In San Francisco data, comparing to the complex-valued neural network, complex-valued GAN further improves classification accuracy than the real-valued neural network, especially for Developed, Low-Density Urban and High-Density Urban with the increase of 44.7\%, 220.9\%, 187.4\%.

\vspace{-0.3cm}
\subsection{Generated Data Analysis}
\label{sec:data analysis}
\vspace{-0.1cm}

In order to analyze the effectiveness of our complex-valued GAN,
we discuss the similarity of actual and generated data in appearance and distribution.
Take Flevoland data for example,
we randomly select 100 pcolors of the real part in diagonal elements of T, as shown in Figure 5. We can clearly find that generated data have high similarity with actual data.
Based on the known data distribution of T matrix\cite{goodman1963statistical}, we further count the distribution of actual and generated data in Figure 6. For actual data, the real and imaginary part statistic histograms of $T_{11}$ shown in (a1) and (a2) and of $T_{12}$ in (a3) and (a4). (b1) - (b4) represent the corresponding statistic histograms of generated $T_{11}$ and $T_{12}$. We can find
the high similarity of generated data with actual data.

\begin{figure}[!htb]
\centering
\setlength{\abovecaptionskip}{-0.0cm}
\setlength{\belowcaptionskip}{-0.6cm}
\includegraphics[width=3.2in]{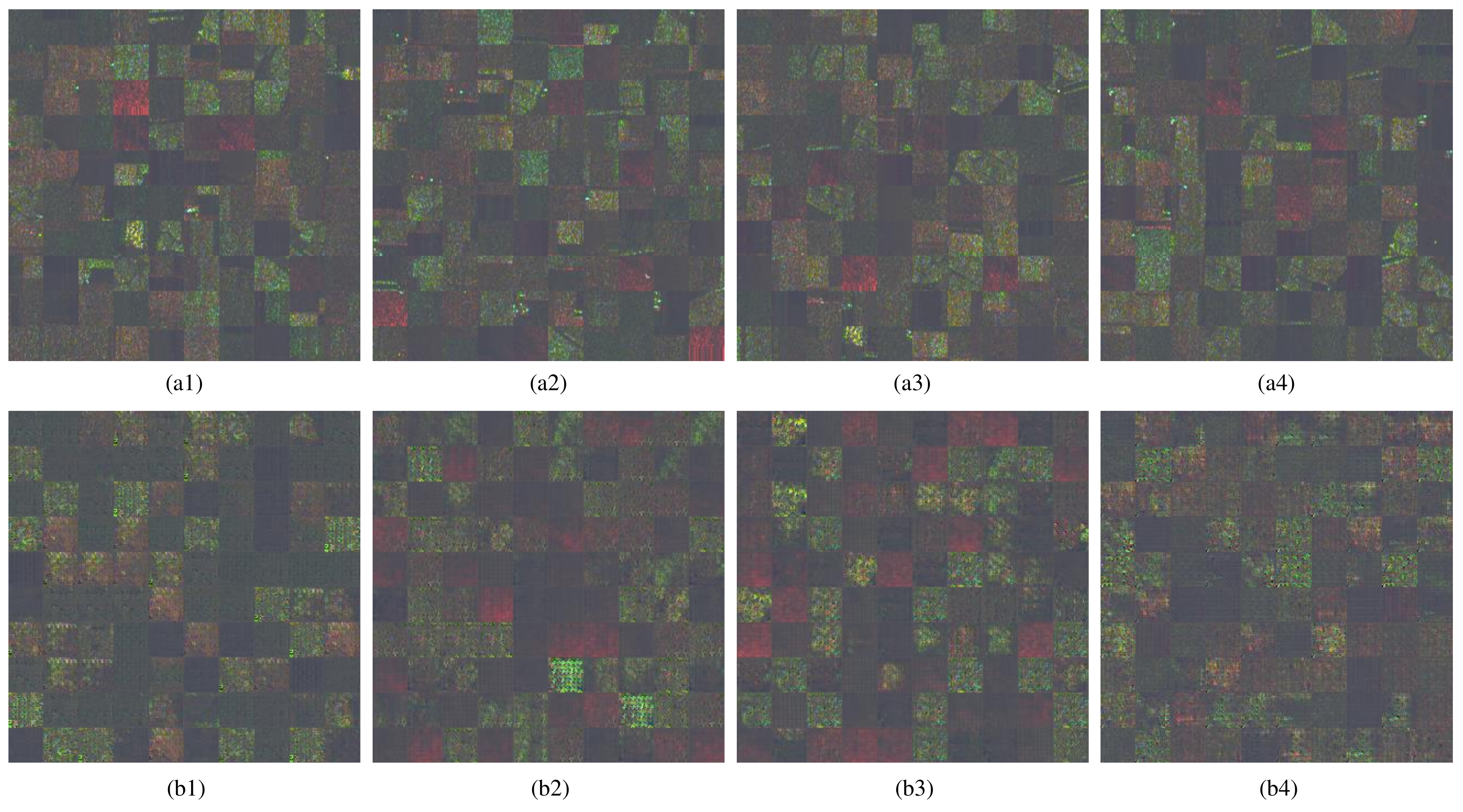}
\caption{\footnotesize{Pcolor comprised by real parts of $T_{11}$, $T_{22}$, $T_{33}$. (a1 - a4) show the actual data image patches. (b1 - b4) show the generated data image patches.}}
\label{fig_University}
\end{figure}

\begin{figure}[!htb]
\centering
\setlength{\abovecaptionskip}{-0.1cm}
\setlength{\belowcaptionskip}{-0.4cm}
\includegraphics[width=3.2in]{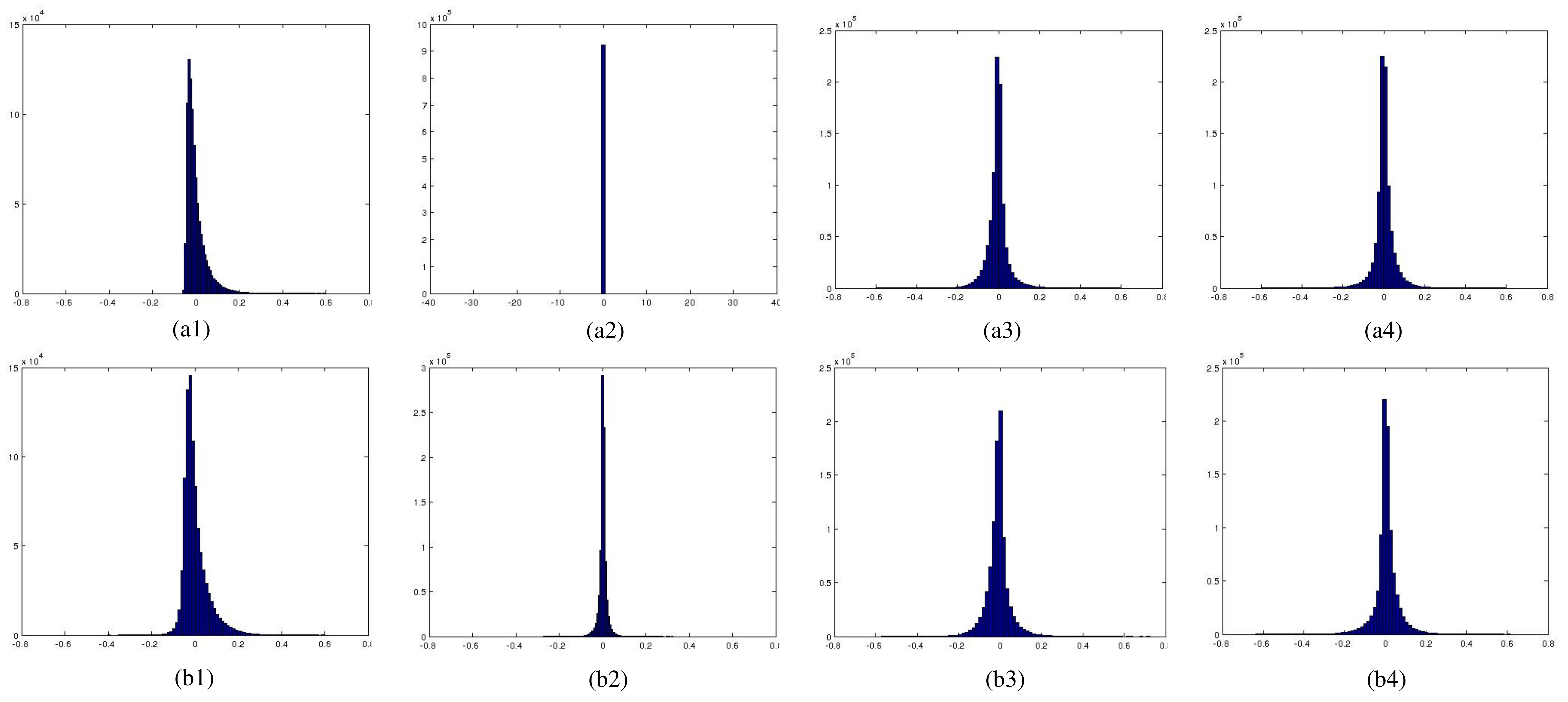}
\caption{\footnotesize{Histograms of representative variables. (a1 - a4) are the statistics of actual data, and (b1 - b4) are the statistics of generated data.}}
\label{fig_University}
\end{figure}

\renewcommand
\arraystretch{1.4}

\vspace{-0.4cm}
\section{CONCLUSION}
\label{sec:con}
\vspace{-0.2cm}

In this paper, a complex-valued GAN is proposed to classify PolSAR data.
Nearly all operations are extended to the complex number field,
and this model obeys the physical meaning of PolSAR data and holds complete phase and amplitude feature.
To the best of our knowledge, this is the first time that complex-valued data is generated by a network,
and the generated data is similar to actual complex-valued data in appearance and distribution
The complex-valued GAN is alternately trained with generated data, labeled data and unlabeled data by semi-supervised learning.
With the utilization of unlabeled and generated samples features, our complex-valued semi-supervised GAN obtains obviously
precede over other models especially when labeled samples are insufficient. It opens up a
new way for our researches on solving the problem of lacking complex-valued samples.

\vspace{-0.15cm}
{\small
\bibliographystyle{IEEEbib}
\addtolength{\itemsep}{-1.3ex} 
\vspace{-0.1cm}
\bibliography{refs}
}
\end{document}